\documentclass[reprint, amsmath,amssymb,aps]{revtex4-1}

\usepackage[utf8]{inputenc} 
\usepackage[T1]{fontenc}    
\usepackage{hyperref}       
\usepackage{url}            
\usepackage{booktabs}       
\usepackage{amsfonts}       
\usepackage{nicefrac}       
\usepackage{microtype}      
\usepackage{bm}
\usepackage{amsmath}
\usepackage{amssymb}
\usepackage{program}
\usepackage{dcolumn}
\usepackage[pdftex]{graphicx}

\newcommand*\diff{\mathop{}\!\mathrm{d}}

\begin{document}

\title{Semiparametric energy-based probabilistic models}

\author{Jan Humplik}
\email{jhumplik@ist.ac.at}
\affiliation{Institute of Science and Technology Austria, Klosterneuburg, Austria}

\author{Gašper Tkačik}
\affiliation{Institute of Science and Technology Austria, Klosterneuburg, Austria}

\begin{abstract}
Probabilistic models can be defined by an energy function, where the probability of each state is proportional to the exponential of the state's negative energy. This paper considers a generalization of energy-based models in which the probability of a state is proportional to an arbitrary positive, strictly decreasing, and twice differentiable function of the state's energy. The precise shape of the nonlinear map from energies to unnormalized probabilities has to be learned from data together with the parameters of the energy function. As a case study we show that the above generalization of a fully visible Boltzmann machine yields an accurate model of neural activity of retinal ganglion cells. We attribute this success to the model's ability to easily capture distributions whose probabilities span a large dynamic range, a possible consequence of latent variables that globally couple the system. Similar features have recently been observed in many datasets, suggesting that our new method has wide applicability.
\end{abstract}

\maketitle

\section{Introduction}

A probabilistic model over a discrete state space is classified as energy-based if it can be written in the form
\begin{equation} \label{energy_based_model}
\begin{aligned}
	p(\mathbf{s};\boldsymbol{\alpha}) &= \frac{e^{-E(\mathbf{s};\boldsymbol{\alpha})}}{Z(\boldsymbol{\alpha})}, \\  
	Z(\boldsymbol{\alpha}) &= \sum_{\mathbf{s}}e^{-E(\mathbf{s};\boldsymbol{\alpha})},
\end{aligned}
\end{equation}
where the \emph{energy} $E(\mathbf{s};\boldsymbol{\alpha})$ is a computationally tractable function of the system's configuration~$\mathbf{s} = (s_1, s_2, \ldots, s_N)$, $\boldsymbol{\alpha}$ is a set of parameters to be learned from data, and $Z(\boldsymbol{\alpha})$ is a normalization constant also known as the \emph{partition function}. Many methods for learning energy-based models exist (e.g. \cite{Tieleman2008,SohlDickstein2011}) which makes them useful in a wide variety of fields, most recently as data analysis tools in neuroscience \cite{Tkacik2014, Koster2014}.  

A popular way of parametrizing the energy function is by decomposing it into a sum of \emph{potentials} representing interactions among different groups of variables, i.e.
\begin{equation} \label{gibbs}
\begin{aligned}
	E(\mathbf{s};\boldsymbol{\alpha}) =& \sum_{i=1}^N \Phi_i(s_i;\boldsymbol{\alpha}_i) + \sum_{i,j=1}^N \Phi_{ij}(s_i, s_j;\boldsymbol{\alpha}_{ij})\ + \\ 
	&+ \sum_{i,j,k=1}^N \Phi_{ijk}(s_i, s_j, s_k;\boldsymbol{\alpha}_{ijk}) + \ldots .
\end{aligned}
\end{equation}
The resulting models, also termed Gibbs random fields \cite{Kindermann1980}, are easy to interpret but, even for moderate~$N$, learning the potential functions from data is intractable unless we can a priori set most of them to zero, or we know how parameters of multiple potentials relate to each other. A common assumption is to consider single- and two-variable potentials only, but many tasks require an efficient parametrization of higher-order interactions.

A powerful way of modeling higher-order dependencies is to assume that they are mediated through hidden variables coupled to the observed system. Many hidden variable models are, however, notoriously hard to learn (e.g. Boltzmann machines \cite{Hinton1986}), and their distributions over observed variables cannot be represented with tractable energy functions. An important exception is the restricted Boltzmann machine (RBM) \cite{Smolensky1986} which is simple to learn even when the dimension of the data is large, and which has proven effective in many applications \cite{Hinton2009, Salakhutdinov2007, Koster2014}. 

This paper considers a new alternative for modeling higher-order interactions. We generalize any model of the form \eqref{energy_based_model} to 
\begin{equation} \label{Venergy_based_model}
\begin{aligned}
	p(\mathbf{s};\boldsymbol{\alpha},V) &= \frac{e^{-V(E(\mathbf{s};\boldsymbol{\alpha}))}}{Z(\boldsymbol{\alpha},V)}, \\  
	Z(\boldsymbol{\alpha}, V) &= \sum_{\mathbf{s}}e^{-V(E(\mathbf{s};\boldsymbol{\alpha}))},
\end{aligned}
\end{equation}
where $V$ is an arbitrary strictly increasing and twice differentiable function which needs to be learned from data together with the parameters $\boldsymbol{\alpha}$. While this defines a new energy-based model with an energy function $E'(\mathbf{s};\boldsymbol{\alpha},V) = V(E(\mathbf{s};\boldsymbol{\alpha}))$, we will keep referring to $E(\mathbf{s};\boldsymbol{\alpha})$ as the energy function. This terminology reflects our interpretation that $E(\mathbf{s};\boldsymbol{\alpha})$ should parametrize local interactions between small groups of variables, e.g. low-order terms in Eq. \eqref{gibbs}, while the function $V$ globally couples the whole system. We will formalize this intuition in Section \ref{why_it_works}. Since setting $V(E) = E$ recovers \eqref{energy_based_model}, we will refer to $V$ simply as \emph{the nonlinearity}. 

Generalized energy-based models have been previously studied in the physics literature on nonextensive statistical mechanics \cite{Hanel20112} but, to our knowledge, they have never been considered as data-driven generative models. If $\mathbf{s}$ is a continuous rather than a discrete vector, then models \eqref{Venergy_based_model} are related to elliptically symmetric distributions \cite{Siwei2009}.

\section{Non-parametric model of the nonlinearity}
\label{nonparametric}
We wish to make as few prior assumptions about the shape of the nonlinearity $V$ as possible. We restrict ourselves to the class of strictly monotone twice differentiable functions for which $V''/V'$ is square-integrable. It is proved in \cite{Ramsay1998} that any such function can be represented in terms of a square-integrable function $W$ and two constants $\gamma_1$ and $\gamma_2$ as
\begin{equation} \label{V_E}
	V(E) = \gamma_1 + \gamma_2 \int_{E_0}^{E} \exp \left( \int_{E_0}^{E'} W(E'') \diff E''  \right) \diff E',
\end{equation}
where $E_0$ is arbitrary and sets the constants to $\gamma_1 = V(E_0)$,  $\gamma_2 = V'(E_0)$. Eq. \eqref{V_E} is a solution to the differential equation $V'' = W V'$, and so $W$ is a measure of the local curvature of $V$. In particular, $V$ is a linear function on any interval on which $W = 0$. 

The advantage of writing the nonlinearity in the form \eqref{V_E} is that we can parametrize it by expanding $W$ in an arbitrary basis without imposing any constraints on the coefficients of the basis vectors. This will allow us to use unconstrained optimization techniques during learning.

We will use piecewise constant functions to parametrize $W$. Let $[E_0,  E_1]$ be an interval containing the range of energies $E(\mathbf{s};\boldsymbol{\alpha})$ which we expect to encounter during learning. We divide the interval $[E_0,  E_1]$ into $Q$ non-overlapping bins of the same width with indicator functions $I_i$, i.e. $I_i(E) = 1$ if $E$ is in the $i$th bin, otherwise $I_i(E) = 0$, and we set $W(E) \equiv W(E;\boldsymbol{\beta}) = \sum_{i=1}^Q \beta_i I_i(E)$. The integrals in Eq.~\eqref{V_E} can be carried out analytically for this choice of $W$ yielding an exact expression for $V$ as a function of $\boldsymbol{\gamma}$ and $\boldsymbol{\beta}$, as well as for its gradient with respect to these parameters (see Appendix A).

The range $[E_0,  E_1]$ and the number of bins $Q$ are metaparameters which potentially depend on the number of samples in the training set, and which should be chosen by cross-validation. 

\section{Maximum likelihood learning}
\label{learning}
Any off-the-shelf method for learning energy-based models is suitable for learning the parameters of the nonlinearity and the energy function simultaneously. The nature of these two sets of parameters is, however, expected to be very different, and an algorithm that takes this fact into account explicitly would likely be useful. As a step in this direction, we present an approximation to the likelihood function of the models \eqref{Venergy_based_model} which can be used to efficiently learn the nonlinearity when the parameters of the energy function are fixed. 

\subsection{The nonlinearity enforces a match between the model and the data probability distributions of the energy}

Let $\rho(E'; \boldsymbol{\alpha}) = \sum_{\mathrm{s}} \delta_{E', E(\mathbf{s};\boldsymbol{\alpha})}$ count the number of states which map to the same energy $E'$. The probability distribution of $E(\mathbf{s};\boldsymbol{\alpha})$ when $\mathbf{s}$ is distributed according to \eqref{Venergy_based_model} is
\begin{equation} \label{p_E_model}
\begin{aligned}
	p(E'; \boldsymbol{\alpha}, V) &= \sum_{\mathbf{s}} p(\mathbf{s};\boldsymbol{\alpha},V)\delta_{E', E(\mathbf{s};\boldsymbol{\alpha})} \\ 
	&= \frac{\rho(E'; \boldsymbol{\alpha})e^{-V(E')}}{Z(\boldsymbol{\alpha},V)}.
\end{aligned}
\end{equation}
Given data $\{\mathbf{s}^{(i)}\}_{i=1}^M$, let $\hat{p}(E'; \boldsymbol{\alpha}) = \frac{1}{M} \sum_{i=1}^M \delta_{E', E(\mathbf{s}^{(i)};\boldsymbol{\alpha})}$ be the data distribution of the energy, and let $\Omega_{\boldsymbol{\alpha}}$ be the image of $E(\mathbf{s};\boldsymbol{\alpha})$. The average log-likelihood of the data can be rewritten as
\begin{equation} \label{likelihood}
\begin{aligned}
	L(\boldsymbol{\alpha}, V) &= -\log Z(\boldsymbol{\alpha}, V) - \sum_{E' \in \Omega_{\boldsymbol{\alpha}}} \hat{p}(E'; \boldsymbol{\alpha})V(E') \\
	&= -\sum_{E' \in \Omega_{\boldsymbol{\alpha}}} \hat{p}(E'; \boldsymbol{\alpha}) \log \rho(E'; \boldsymbol{\alpha}) \\
	&\qquad + \sum_{E' \in \Omega_{\boldsymbol{\alpha}}} \hat{p}(E'; \boldsymbol{\alpha}) \log p(E'; \boldsymbol{\alpha}, V),
\end{aligned}
\end{equation}
where the first line is a standard expression for energy based models, and the second line follows by substituting the logarithm of \eqref{p_E_model}.

Eq. \eqref{likelihood} has a simple interpretation. The last term, which is the only one depending on $V$, is the average log-likelihood of the samples $\{E(\mathbf{s}^{(i)};\boldsymbol{\alpha})\}_{i=1}^M$ under the model $p(E; \boldsymbol{\alpha}, V)$, and so, for any $\boldsymbol{\alpha}$, the purpose of the nonlinearity is to reproduce the data probability distribution of the energy. 

Our restriction that $V$ is a twice differentiable increasing function can be seen as a way of regularizing learning. If $V$ was arbitrary, then, for any $\boldsymbol{\alpha}$, an upper bound on the likelihood \eqref{likelihood} is attained when $\hat{p}(E; \boldsymbol{\alpha}) = p(E; \boldsymbol{\alpha}, V)$. According to \eqref{p_E_model}, this can be satisfied with any function $V$ (potentially infinite at some points) such that for all $E \in \Omega_{\boldsymbol{\alpha}}$
\begin{equation} \label{eq_exact}
	V(E) = \log \rho(E; \boldsymbol{\alpha}) - \log \hat{p}(E; \boldsymbol{\alpha}) + \text{const.}\end{equation}
Energy functions often assign distinct energies to distinct states in which case the choice \eqref{eq_exact} leads to a model which exactly reproduces the empirical distribution of data, and hence overfits. 

\subsection{Integral approximation of the partition function, and of the likelihood}
\label{sec_approx}

The partition function \eqref{Venergy_based_model} can be rewritten as
\begin{equation}
	Z(\boldsymbol{\alpha}, V) = \sum_{E' \in \Omega_{\boldsymbol{\alpha}}} \rho(E'; \boldsymbol{\alpha}) e^{-V(E')}.
\end{equation}
The size of $\Omega_{\boldsymbol{\alpha}}$ is generically exponential in the dimension of $\mathbf{s}$, making the above sum intractable for large systems. To make it tractable, we will use a standard trick from statistical mechanics, and approximate the sum over distinct energies by smoothing $\rho(E; \boldsymbol{\alpha})$. 

Let $[E_{\text{min}},  E_{\text{max}}]$ be an interval which contains the set $\Omega_{\mathbf{\alpha}}$, and which is divided into $K$ non-overlapping bins~$I_i \in [E_{\text{min}},  E_{\text{max}}]$ of width $\Delta = (E_{\text{max}}-E_{\text{min}})/K$. Let $E_i$ be the energy in the middle of the $i$th bin. The partition function can be approximated as
\begin{equation} \label{Z_approx}
\begin{aligned}
	Z(\boldsymbol{\alpha}, V) &= \sum_{i=1}^K \sum_{E' \in I_i} \rho(E'; \boldsymbol{\alpha}) e^{-V(E')} \\
	&= \sum_{i=1}^K e^{-V(E_i)} \sum_{E' \in I_i} \rho(E'; \boldsymbol{\alpha}) + o(\Delta) \\
	&\equiv \sum_{i=1}^K e^{-V(E_i)} \bar{\rho}(E_i; \boldsymbol{\alpha}) \Delta + o(\Delta),
\end{aligned}	
\end{equation}
where $\bar{\rho}(E_i; \boldsymbol{\alpha}) = (1/\Delta) \sum_{E' \in I_i} \rho(E'; \boldsymbol{\alpha})$ is the \emph{density of states}, and $o(\Delta) \to 0$ as $\Delta \to 0$ since $V$ is differentiable. 

The approximation $o(\Delta) \approx 0$ is useful for two reasons. First, in most problems, this approximation becomes very good already when the number of bins $K$ is still much smaller than the size of $\Omega_{\boldsymbol{\alpha}}$. Second, an efficient Monte Carlo method, the Wang and Landau algorithm \cite{Wang2001, Belardinelli2007}, exists for estimating the density of states for any fixed value of $\boldsymbol{\alpha}$. 

The approximation \eqref{Z_approx} yields the following approximation of the average log-likelihood \eqref{likelihood}
\begin{equation} \label{L_approx}
\begin{aligned}
	L(\boldsymbol{\alpha}, V) \approx & -\log \left( \sum_{i=1}^K e^{-V(E_i)} \bar{\rho}(E_i; \boldsymbol{\alpha}) \Delta \right) \\ 
	&- \frac{1}{M}\sum_{i=1}^M V(E(\mathbf{s}^{(i)};\boldsymbol{\alpha})),
\end{aligned}
\end{equation}
which can be maximized with respect to the parameters of $V$ using any gradient-based optimization technique (after specifying the metaparameters $E_{min}$, $E_{max}$, and $K$). 

Many learning algorithms represent information about the model $p(\mathbf{s};\boldsymbol{\alpha},V)$ as a finite list of samples from the model. This representation is necessarily bad at capturing the low probability regions of $p(E; \boldsymbol{\alpha}, V)$. According to \eqref{likelihood}, this means that any such algorithm is expected to be inefficient for learning $V$ in these regions. On the other hand, the density of states $\bar{\rho}$ estimated with the Wang and Landau algorithm carries information about the low probability regions of $p(E; \boldsymbol{\alpha}, V)$ with the same precision as about the high probability regions, and so algorithms based on maximizing \eqref{L_approx} should be efficient at learning the nonlinearity.

The approximation \eqref{L_approx} cannot be used to learn the parameters $\boldsymbol{\alpha}$ of the energy function, and so the above algorithm has to be supplemented with other techniques. 

\section{Experiments: modeling neural activity of retinal ganglion cells}
\label{neurons}

A major question in neuroscience is how populations of neurons, rather than individual cells, respond to inputs from the environment. It is well documented that single neuron responses are correlated with each other but the precise structure of the underlying redundancy, and its functional role in information processing is only beginning to be unraveled \cite{Tkacik2014}. 

The response of a population of $N$ neurons during a short time window can be represented as a binary vector $\mathbf{s} \in \{0,1\}^N$ by assigning a $1$ to every neuron which elicited a spike. We pool all population responses recorded during an experiment, and we ask what probabilistic model would generate these samples. This question was first asked in a seminal paper \cite{Schneidman2006} which showed that, for small networks of retinal neurons ($N < 20$), the fully visible Boltzmann machine, or a \emph{pairwise model},
\begin{equation} \label{pairwise}
	p(\mathbf{s};\boldsymbol{J}) = \frac{1}{Z(\boldsymbol{J})}\exp \left(-\sum_{i,j=1}^N J_{ij} s_i s_j\right),
\end{equation}  
is a good description of the data. Later it was realized that for large networks ($40 < N < 120$) pairwise models cannot accurately capture probability distributions of data statistics which average across the whole population such as the total population activity~$K(\mathbf{s}) = \sum_{i=1}^N s_i$. This issue was solved by the introduction of the so-called \emph{K-pairwise models} \cite{Tkacik2014},
\begin{equation}
\begin{aligned}
	p(\mathbf{s};\boldsymbol{J}, \boldsymbol{\phi}) = \frac{1}{Z(\boldsymbol{J}, \boldsymbol{\phi})}& \exp \left(-\sum_{i,j=1}^N J_{ij} s_i s_j \right. \\ 
	&- \left. \sum_{k=0}^N \phi_k \delta_{k,K(\mathbf{s})}\right).
\end{aligned}
\end{equation}  

Here we look at the performance of two additional models. A \emph{semiparametric pairwise model},
which is a generalization \eqref{Venergy_based_model} of the pairwise model, and a restricted Boltzmann machine with $N/2$ hidden units which is known to be an excellent model of higher order dependencies. 

The philosophy of our comparison is not to find a state-of-the art model, but rather to contrast several different models of comparable complexity (measured as the number of parameters) in order to demonstrate that the addition of a simple nonlinearity to an energy-based model can result in a significantly better fit to data.

\subsection{Data}
We analyze a simultaneous recording from $160$ neurons in a salamander's retina which is presented with $297$ repetitions of a $19$ second natural movie. The data was collected as part of \cite{Tkacik2014}, and it contains a total of $\sim 2.8 \times 10^5$ population responses.

Our goal is to compare the performance, measured as the out-of-sample log-likelihood per sample per neuron, of the above models across several subnetworks of different sizes. To this end, we use our data to construct $48$ smaller datasets as follows. We randomly select $40$ neurons from the total of $160$ as the first dataset. Then we augment this dataset with $20$ additional neurons to yield the second dataset, and we keep repeating this process until we have a dataset of $140$ neurons. This whole process is repeated $8$ times, resulting in $8$ datasets for each of the $6$ different network sizes. Only the datasets with $40$ and $60$ neurons do not have significant overlaps which is a limitation set by the currently available experiments. For each dataset, we set aside responses corresponding to randomly selected $60$ (out of 297) repetitions of the movie, and use these as test data.

\subsection{Results}

\begin{figure}
	\centering
	\includegraphics[width=\linewidth]{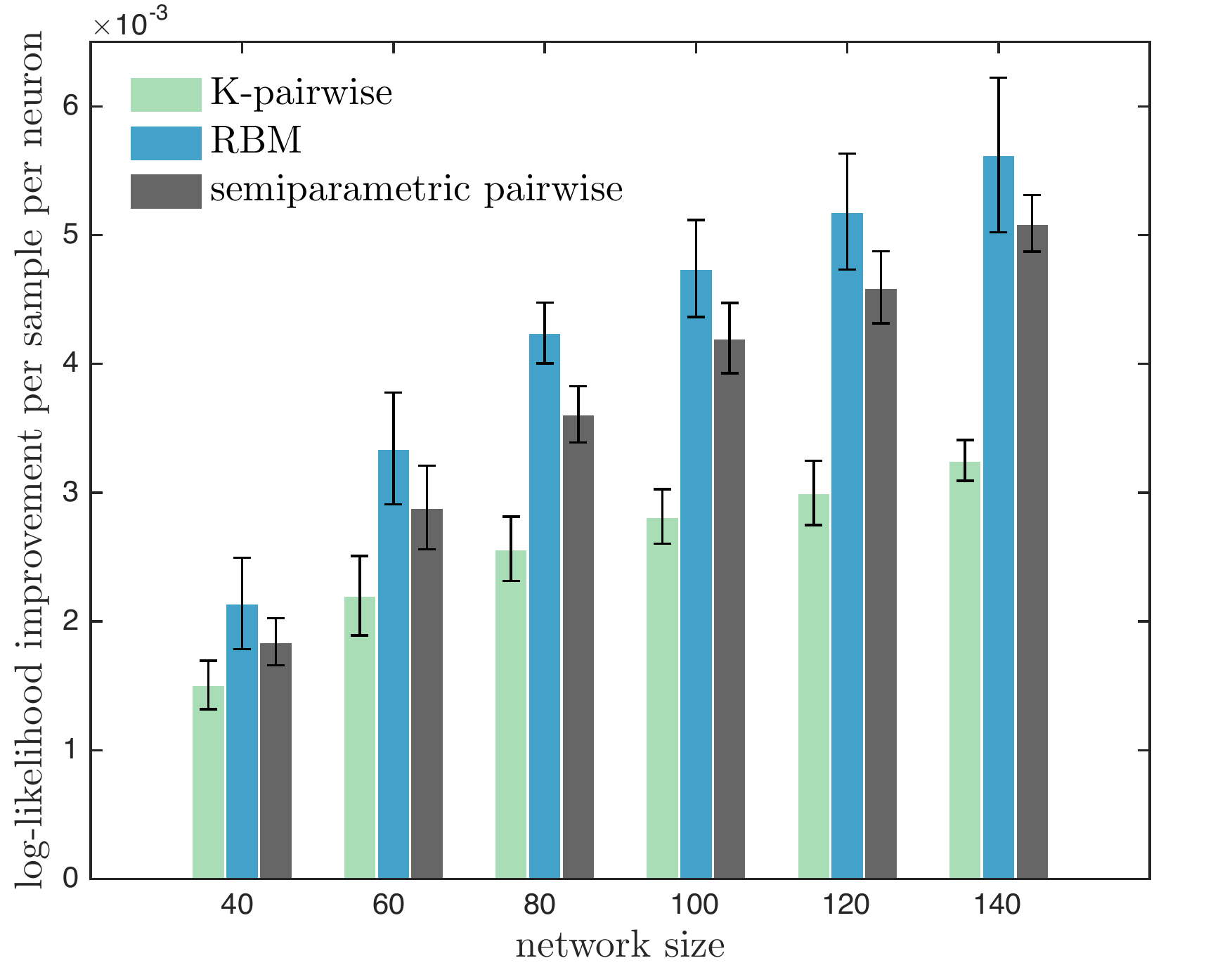}
	\caption{Out-of-sample log-likelihood improvement per sample per neuron averaged over subnetworks. Errorbars denote variation over subnetworks (one standard deviation), not the accuracy of likelihood estimates. Baseline is the pairwise model.}
	\label{fig1}
\end{figure}

Our results are summarized in Figure \ref{fig1}. We see that both semiparametric pairwise models and RBMs with $N/2$ hidden units substantially outperform the previously considered K-pairwise models. In particular, the improvement increases as we go to larger networks. RBMs are consistently better than semiparametric pairwise models, but, interestingly, this gap does not seem to scale with the network size. 

\begin{figure} \label{fig_nonlinearity}
	\centering
	\includegraphics[width=\linewidth]{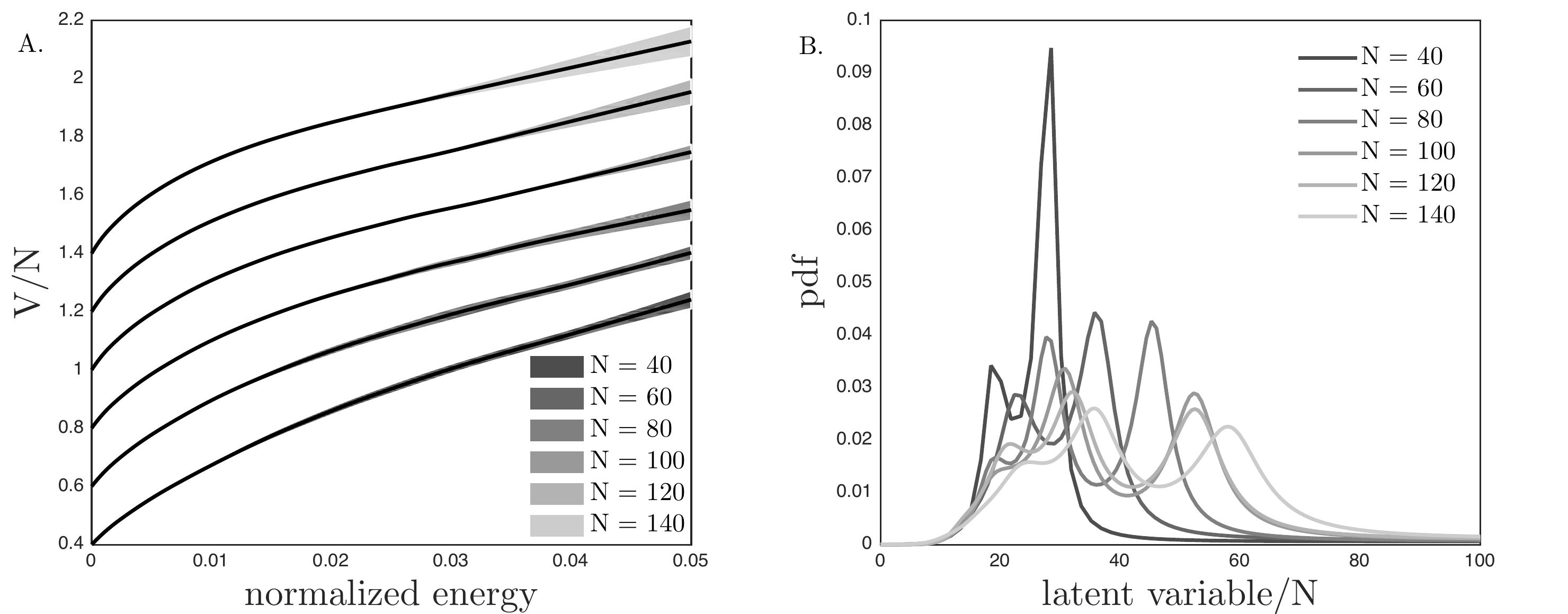}
	\caption{A. Inferred nonlinearities of the semiparametric pairwise model. The shift along the y-axis is arbitrary, and its only purpose is to increase readability. Black lines are averages over subnetworks. Shaded regions denote variation across subnetworks (one standard deviation). Curves are ordered in increasing order (bottom is $N=40$, top is $N=140$). B. Inferred probability densities of the latent variable for one sequence of subnetworks.}
	\label{fig2}
\end{figure}

The inferred nonlinearity $V$ for semiparametric pairwise models does not vary much across subnetworks of the same size. This is not surprising for large networks since there is a substantial overlap between the datasets, but it is nontrivial for smaller networks. The average inferred nonlinearities are shown in Figure \ref{fig2}A. Semiparametric pairwise models have one free parameter since the transformation $\boldsymbol{J} \to \kappa \boldsymbol{J}, V(E) \to V(E/\kappa)$ does not change their probability distribution. Therefore, to make the comparison of $V$ across different datasets fair, we fix $\kappa$ by requiring $\sum_{i,j} J_{ij}^2 = 1$. Furthermore, the nonlinearities in Figure \ref{fig2}A are normalized by $N$ because we expect the probabilities in a system of dimension $N$ to scale as $e^{-N}$. The inferred nonlinearities are approximately concave, and their curvature increases with the network size. 

\subsection{Training the models}
Training was done by a variation of Persistent Contrastive Divergence \cite{Tieleman2008} which performs an approximate gradient ascent on the log-likelihood of any energy-based model \eqref{energy_based_model}. Given an initial guess of the parameters $\boldsymbol{\alpha}_0$, and a list of $M_s$ samples drawn from $p(\mathbf{s};\boldsymbol{\alpha}_0)$, the algorithm can be summarized as
\begin{program}
\FOR t:=1 \TO L
	\boldsymbol{\alpha}_t = \boldsymbol{\alpha}_{t-1} + \eta (\mathbf{E}[\nabla_{\boldsymbol{\alpha}}E(\mathbf{s};\boldsymbol{\alpha}_{t-1})]_{\text{samples}_{t-1}} 
	\qquad \qquad \qquad \quad - \mathbf{E}[\nabla_{\boldsymbol{\alpha}}E(\mathbf{s};\boldsymbol{\alpha}_{t-1})]_{\text{data}})
	\text{samples}_t = \text{GIBBS}^n(\text{samples}_{t-1},\boldsymbol{\alpha}_t)
\end{program}
where $L$ is the number of iterations, $\eta$ is the learning rate, $\mathbf{E}[\cdot]_{\text{list}}$ denotes an average over the list of states, and $\text{GIBBS}^n$ represents $n$ applications (i.e. $n$ neurons are switched) of the Gibbs sampling transition operator. 

In the case of semiparametric pairwise models, Persistent Contrastive Divergence was used as a part of an alternating maximization algorithm in which we learned the nonlinearity by maximizing the approximate likelihood  \eqref{L_approx} while keeping the couplings $\boldsymbol{J}$ fixed, and then learned the couplings using Persistent Contrastive Divergence with the nonlinearity fixed. Details of the learning algorithms for all models are described in Appendix B. 

The most interesting metaparameter is the number of bins $Q$ necessary to model the nonlinearity $V$. We settled on $Q=12$ but we observed that decreasing it to $Q=6$ would not significantly change the training likelihood. However, $Q=12$ yielded more consistent nonlinearities over different subgroups. 

\subsection{Estimating likelihood and controlling for overfitting}

Estimating likelihood of our data is easy because the state $\mathbf{s}_0 = (0,\ldots,0)$ occurs with high probability ($\sim 0.2$), and all the inferred models retain this property. Therefore, for any energy-based model, we estimated $p(\mathbf{s}_0;\boldsymbol{\alpha})$ by drawing $3\times 10^6$ samples using Gibbs sampling, calculated the partition function as $Z(\boldsymbol{\alpha}) = \exp (-E(\mathbf{s}_0); \boldsymbol{\alpha})/p(\mathbf{s}_0;\boldsymbol{\alpha})$, and used it to calculate the likelihood.

The models do not have any explicit regularization. We tried to add a smoothed version of L1 regularization on coupling matrices but we did not see any improvement in generalization using a cross-validation on one of the training datasets. Certain amount of regularization is due to sampling noise in the estimates of likelihood gradients, helping us to avoid overfitting.

\section{Why does it work?}
\label{why_it_works}

Perhaps surprisingly, the addition of a simple nonlinearity to the pairwise energy function $E(\mathbf{s};\boldsymbol{J}) = \sum_{i,j=1}^N J_{ij} s_i s_j$ significantly improves the fit to data. Here we give heuristic arguments that this should be expected whenever the underlying system is globally coupled. 

\subsection{Many large complex systems are globally coupled}
Let $p_N(\mathbf{s})$ be a sequence of positive probabilistic models ($\mathbf{s}$ is of dimension $N$), and suppose that $p_N(\mathbf{s})$ can be (asymptotically) factorized into subsystems statistically independent of each other whose number is proportional to $N$. Then $(1/N)\log p_N(\mathbf{s})$ is an average of independent random variables, and we expect its standard deviation $\sigma(\log p_N(\mathbf{s}))/N$ to vanish in the $N \to \infty$ limit. Alternatively, if $\sigma(\log p_N(\mathbf{s}))/N \nrightarrow 0$, then the system cannot be decomposed into independent subsystems, and there must be some mechanism globally coupling the system together.

It has been argued that many natural systems \cite{Mora2011} including luminance in natural images \cite{Stephens2013}, amino acid sequences of proteins \cite{Mora2010}, and neural activity such as the one studied in Sec. \ref{neurons} \cite{Tkacik2015, Mora2015} belong to the class of models whose log-probabilities per dimension have large variance even though their dimensionality is big. Therefore, models of such systems should reflect the prior expectation that there is a mechanism which couples the whole system together. In our case, this mechanism is the nonlinearity $V$.
\subsection{Mapping the nonlinearity to a latent variable}

Recent work attributes the strong coupling observed in many systems to the presence of latent variables (\cite{Schwab2014, Aitchison2014}). We can rewrite the model \eqref{Venergy_based_model} in terms of a latent variable considered in \cite{Schwab2014} if we assume that $\exp (-V(E))$ is a totally monotone function, i.e. that it is continuous for $E \geq 0$ (we assume, without loss of generality, that $E(\mathbf{s};\boldsymbol{\alpha}) \geq 0$), infinitely differentiable for $E > 0$, and that $(-1)^n d^n \exp (-V(E))/dE^n \geq 0$ for $n \geq 0$. Bernstein's theorem \cite{Widder1946} then asserts that we can rewrite the model \eqref{Venergy_based_model} as
\begin{equation}  \label{eq_latent}
\begin{aligned}
	\frac{e^{-V(E(\mathbf{s};\boldsymbol{\alpha}))}}{Z(\boldsymbol{\alpha},V)} &= \int_{0}^{\infty} q(h)\frac{e^{-hE(\mathbf{s}; \boldsymbol{\alpha})}}{Z(h;\boldsymbol{\alpha})} \diff h, \\ 
	Z(h;\boldsymbol{\alpha}) &= \sum_{\mathbf{s}} e^{-hE(\mathbf{s};\boldsymbol{\alpha})},
\end{aligned}
\end{equation}
where $q(h)$ is a probability density (possibly containing delta functions). Suppose that the energy function has the form \eqref{gibbs}. Then we can interpret \eqref{eq_latent} as a latent variable $h$ being coupled to every group of interacting variables, and hence inducing a coupling between the whole system whose strength depends on the size of the fluctuations of $h$. 
 
While the class of positive, twice differentiable, and decreasing functions that we consider is more general than the class of totally monotone functions, we can find the maximum likelihood densities $q(h)$ which correspond to the pairwise energy functions inferred in Section \ref{neurons} using the semiparametric pairwise model. We model $q(h)$ as a histogram, and maximize the likelihood under the model \eqref{eq_latent} by estimating $Z(h;\boldsymbol{J})$ using the approximation \eqref{Z_approx}. The maximum likelihood densities are shown in Figure \ref{fig2}B for one particular sequence of networks of increasing size. The units of the latent variables are arbitrary, and set by the scale of $\boldsymbol{J}$ which we normalize so that $\sum_{i,j} J_{ij}^2 = 1$. The bimodal structure of the latent variables is observed across all datasets. We do not observe a significant decrease in likelihood by replacing the nonlinearity $V$ with the integral form \eqref{eq_latent}. Therefore, at least for the data in Sec. \ref{neurons}, the nonlinearity can be interpreted as a latent variable globally coupling the system. 

\subsection{Asymptotic form of the nonlinearity}

Suppose that the true system $\hat{p}_N(\mathbf{s})$ which we want to model with \eqref{Venergy_based_model} satisfies $\sigma(\log p_N(\mathbf{s}))/N \nrightarrow 0$. Then we also expect that energy functions $E_N(\mathbf{s};\boldsymbol{\alpha})$ which accurately model the system satisfy $\sigma(E_N(\mathbf{s};\boldsymbol{\alpha}))/N \nrightarrow 0$. In the limit $N \to \infty$, the approximation of the likelihood \eqref{L_approx} becomes exact when $V$ is differentiable ($\Delta$ has to be appropriately scaled with $N$), and arguments which led to \eqref{eq_exact} can be reformulated in this limit to yield
\begin{equation} \label{eq_limit}
	V_N(E) = \log \bar{\rho}_N(E; \boldsymbol{\alpha}) - \log \hat{\bar{p}}_N(E; \boldsymbol{\alpha}) + \text{const.},
\end{equation}
where $\bar{\rho}_N(E; \boldsymbol{\alpha})$ is the density of states, and $\hat{\bar{p}}_N(E; \boldsymbol{\alpha})$ is now the probability density of $E_N(\mathbf{s};\boldsymbol{\alpha})$ under the true model. In statistical mechanics, the first term $\log \bar{\rho}_N(E; \boldsymbol{\alpha})$ is termed the \emph{microcanonical entropy}, and is expected to scale linearly with $N$. On the other hand, because $\sigma(E_N(\mathbf{s};\boldsymbol{\alpha}))/N \nrightarrow 0$, we expect the second term to scale at most as $\log N$. Thus we make the prediction that if the underlying system cannot be decomposed into independent subsystems, then the maximum likelihood nonlinearity satisfies $V(E) \approx \log \bar{\rho}(E; \boldsymbol{\alpha})$ up to an arbitrary constant.

\begin{figure}
	\centering
	\includegraphics[width=\linewidth]{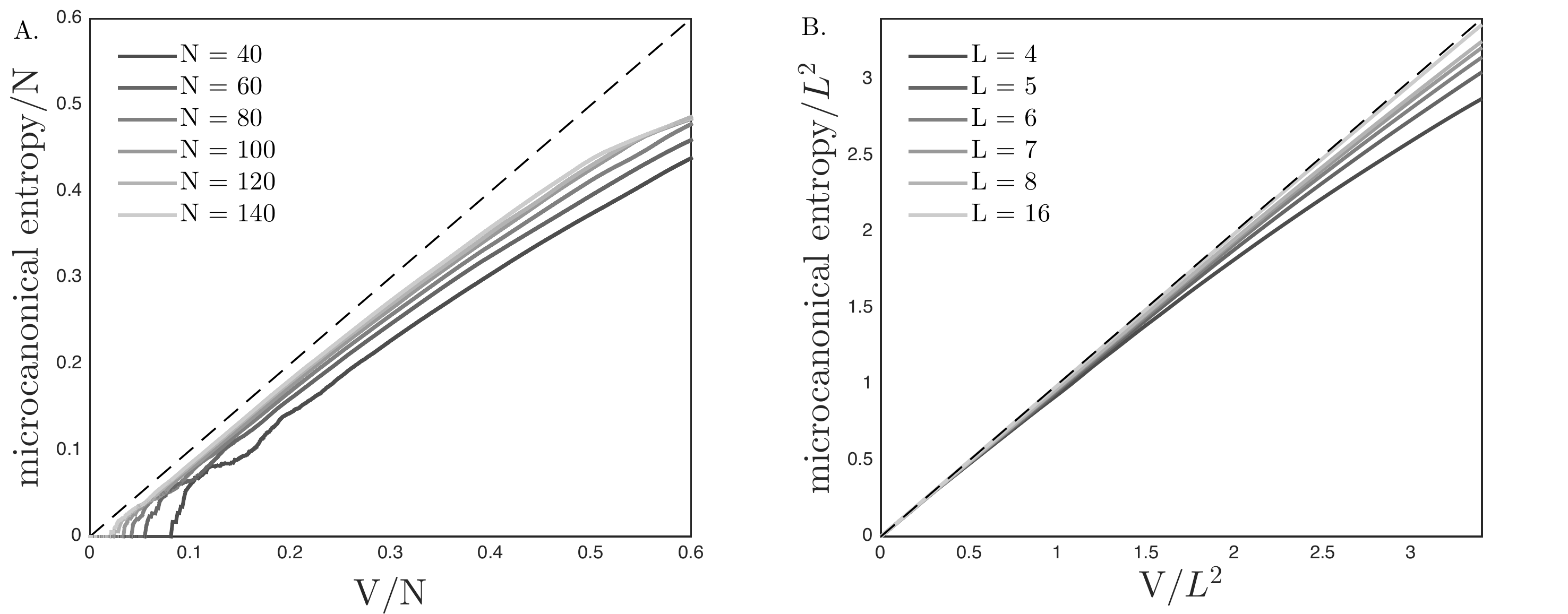}
	\caption{A. Plot of the inferred nonlinearity vs the microcanonical entropy for the semiparametric pairwise model. B. The same for the elliptically symmetrical model of luminance distribution in natural images.}
	\label{fig3}
\end{figure}

In Figure \ref{fig3}A we show a scatter plot of the inferred nonlinearity for one sequence of subnetworks in Sec. \ref{neurons} vs the microcanonical entropy estimated using the Wang and Landau algorithm. While the convergence is slow, the plot suggests that these functions approach each other as the network size increases. To demonstrate this prediction on yet another system, we used the approach in \cite{Siwei2009} to fit the semiparametric pairwise model with $\mathbf{s} \in \mathbb{R}^{L^2}$ to $L \times L$ patches of pixel log-luminances in natural scenes from the database \cite{Tkacik2011}. This model has an analytically tractable density of states which makes the inference simple. Figure \ref{fig3}B shows the relationship between the inferred nonlinearity and the microcanonical entropy for collections of patches which increase in size, confirming our prediction that the nonlinearity should be given by the microcanonical entropy.

\section{Conclusions}

We presented a tractable extension of any energy-based model which can be interpreted as augmenting the original model with a latent variable. As demonstrated on the retinal activity data, this extension can yield a substantially better fit to data even though the number of additional parameters is negligible compared to the number of parameters of the original model. In light of our results, we hypothesize that combing a nonlinearity with the energy function of a restricted Boltzmann machine might yield a model of retinal activity which is not only accurate, but also simple as measured by the number of parameters. Simplicity is an important factor in neuroscience because of experimental limitations on the number of samples. We plan to pursue this hypothesis in future work.

Our models are expected to be useful whenever the underlying system cannot be decomposed into independent components. This phenomenon has been observed in many natural systems, and the origins of this global coupling, and especially its analogy to physical systems at critical points, have been hotly debated. Our models effectively incorporate the prior expectations of a global coupling in a simple nonlinearity, making them superior to models based on Gibbs random fields which might need a large number of parameters to capture the same dependency structure.

\section*{Acknowledgments}
We thank David Schwab, Elad Schneidman, and William Bialek for helpful discussions. This work was supported in part by HFSP Program Grant RGP0065/2012 and Austrian Science Fund (FWF) grant P25651.

\appendix
\onecolumngrid
\section{Exact expression for the nonlinarity and its gradient}

Let $W(E;\boldsymbol{\beta}) = \sum_{i=1}^Q \beta_i I_i(E)$ where $I_i$ are the indicator functions defined in the main text. For $E < E_0$, we have $V(E;\boldsymbol{\gamma}, \boldsymbol{\beta}) = \gamma_1 + \gamma_2(E-E_0)$. For $E > E_0$ we have $V(E;\boldsymbol{\gamma}, \boldsymbol{\beta}) = \gamma_1 + \gamma_2f(E;\boldsymbol{\beta})$, where
\begin{equation}
\begin{aligned}
	 f(E;\boldsymbol{\beta}) &= \int_{E_0}^{E} \exp \left( \int_{E_0}^{E'} W(E'';\boldsymbol{\beta}) \diff E''  \right) \diff E' \\
	&= \sum_{i=1}^{[E]-1} \exp\left(\Delta \sum_{j=1}^{i-1} \beta_j \right) \frac{\exp(\Delta \beta_i)-1}{\beta_i}+ \exp \left(\Delta \sum_{j=1}^{[E]-1} \beta_j \right)  \frac{\exp(\beta_{[E]}(E - ([E]-1)\Delta) )-1}{\beta_{[E]}}.
\end{aligned}
\end{equation}
We define $[E]$ as the number of the bin that contains $E$. If $E > E_1$, then we define $[E] = Q + 1$, and $\beta_{Q+1} = 0$. 

The gradient is
\begin{equation}
	\frac{\partial V(E;\boldsymbol{\gamma}, \boldsymbol{\beta})}{\partial \gamma_1} = 1,
\end{equation}
\begin{equation}
	\frac{\partial V(E;\boldsymbol{\gamma}, \boldsymbol{\beta})}{\partial \gamma_2} = f(E;\boldsymbol{\beta}),
\end{equation}
\begin{equation}
	\frac{\partial V(E;\boldsymbol{\gamma}, \boldsymbol{\beta})}{\partial \beta_k} = \gamma_2 \frac{f(E;\boldsymbol{\beta})}{\partial \beta_k}.
\end{equation}
If $k > [E]$, then
\begin{equation}
	\frac{\partial f(E;\boldsymbol{\beta})}{\partial \beta_k} = 0.
\end{equation}
If $k = [E]$, then
\begin{equation}
	\frac{\partial f(E;\boldsymbol{\beta})}{\partial \beta_k} = \exp\left(\Delta \sum_{j=1}^{[E]-1} \beta_j \right)\frac{\exp (\Delta \beta_{[E]})\Delta \beta_{[E]} - \exp (\Delta \beta_{[E]}) + 1}{\beta_{[E]}^2}.
\end{equation}
If $k < [E]$, then
\begin{equation}
\begin{aligned}
	\frac{\partial f(E;\boldsymbol{\beta})}{\partial \beta_k} &= \exp\left(\Delta \sum_{j=1}^{k-1} \beta_j \right)\frac{\exp (\Delta \beta_{k})\Delta \beta_{k} - \exp (\Delta \beta_{k}) + 1}{\beta_{k}^2} + \Delta \sum_{i=k+1}^{[E]-1} \exp\left(\Delta \sum_{j=1}^{i-1} \beta_j \right) \frac{\exp(\Delta \beta_i)-1}{\beta_i} \\
	&\qquad + \Delta \exp\left(\Delta \sum_{j=1}^{[E]-1} \beta_j \right) \frac{\exp(\beta_{[E]})(E - ([E]-1)\Delta)-1}{\beta_{[E]}}.
\end{aligned}
\end{equation}

\section{Details of training}
\subsection{Pairwise models, K-pairwise models, and RBMs}
Pairwise models, K-pairwise models, and RBMs were all trained using Persistent Contrastive Divergence with $\eta = 2$, $n = 2N$, and with initial parameters drawn from a normal distribution with $0$ mean and $0.1$ standard deviation. We iterated the algorithm three times, first with $L = 1000, M_s = 3\times 10^4$, then with $L = 1000, M_s = 3\times 10^5$, and finally with $L = 500, M_s = 3\times 10^6$ (the last step had $L = 1000$ for K-pairwise models, and also for RBMs when $N=140$). 

\subsection{Semiparametric pairwise models}
We initialized the coupling matrix $\boldsymbol{J}_0$ as the one learned using a pairwise model. The $E_0$ and $E_1$ metaparameters of $V$ were set to the minimum and maximum energy $E(\mathbf{s};\boldsymbol{J}_0) = \sum_{ij} J_{0ij}s_i s_j$ observed in the training set. We set $Q = 6$, and we initialized the parameters $\boldsymbol{\gamma}$ and $\boldsymbol{\beta}$ of the nonlinearity by maximizing the approximate likelihood with $\boldsymbol{J} = \boldsymbol{J}_0$ fixed. The metaparameters $E_{\text{min}}$ and $E_{\text{max}}$ for the approximate likelihood were set to the minimum, and twice the maximum of $E(\mathbf{s};\boldsymbol{J}))$ over the training set. $K$ was set between $2000$ and $3000$. The density of states was estimated with a variation of the algorithm described in \cite{Belardinelli2007} with accuracy $F_{\text{final}} = 10^{-8}$.

Starting with these initial parameters, we ran two iterations of Persistent Contrastive Divergence with $n=2N$, simultaneously learning the coupling matrix and the nonlinearity. The first iteration had $L = 1000, M_s = 3\times 10^4$, and the second one $L = 1000, M_s = 3\times 10^5$. In order for the learning to be stable, we had to choose different learning rates for the coupling matrix ($\eta = 2$), and for $\boldsymbol{\gamma}$ and $\boldsymbol{\beta}$ ($\eta = 10^{-4}$).

For the last step, we adjusted the metaparameters of the nonlinearity so that $E_0$ and $E_1$ are the minimum and maximum observed energies with the current $\boldsymbol{J}$, and $Q=12$. We maximized the approximate likelihood to infer the nonlinearity with these new metaparameters. Then we fixed $V$, and ran a Persistent Contrastive Divergence ($\eta = 2, n = 2N, L = 500, M_s = 3\times 10^6$) learning $\boldsymbol{J}$. Finally we maximized the approximate likelihood with fixed $\boldsymbol{J}$ to get the final nonlinearity. 

\twocolumngrid


\begin{thebibliography}{10}

\bibitem{Tieleman2008}
Tijmen Tieleman.
\newblock Training restricted boltzmann machines using approximations to the
  likelihood gradient.
\newblock In {\em Proceedings of the 25th International Conference on Machine
  Learning}, ICML '08, pages 1064--1071, New York, NY, USA, 2008. ACM.

\bibitem{SohlDickstein2011}
Jascha Sohl-Dickstein, Peter~B. Battaglino, and Michael~R. DeWeese.
\newblock New method for parameter estimation in probabilistic models: Minimum
  probability flow.
\newblock {\em Phys. Rev. Lett.}, 107:220601, Nov 2011.

\bibitem{Tkacik2014}
Gašper Tkačik, Olivier Marre, Dario Amodei, Elad Schneidman, William Bialek,
  and Michael~J. Berry, II.
\newblock Searching for collective behavior in a large network of sensory
  neurons.
\newblock {\em PLoS Comput Biol}, 10(1):e1003408, 01 2014.

\bibitem{Koster2014}
Urs Köster, Jascha Sohl-Dickstein, Charles~M. Gray, and Bruno~A. Olshausen.
\newblock Modeling higher-order correlations within cortical microcolumns.
\newblock {\em PLoS Comput Biol}, 10(7):e1003684, 07 2014.

\bibitem{Kindermann1980}
R.~Kindermann and J.~L. Snell.
\newblock {\em Markov Random Fields and Their Applications}.
\newblock AMS books online. American Mathematical Society, 1980.

\bibitem{Hinton1986}
G.~E. Hinton and T.~J. Sejnowski.
\newblock {Parallel Distributed Processing: Explorations in the Microstructure
  of Cognition, Vol. 1}.
\newblock chapter Learning and Relearning in Boltzmann Machines, pages
  282--317. MIT Press, Cambridge, MA, USA, 1986.

\bibitem{Smolensky1986}
P.~Smolensky.
\newblock {Parallel Distributed Processing: Explorations in the Microstructure
  of Cognition, Vol. 1}.
\newblock chapter Information Processing in Dynamical Systems: Foundations of
  Harmony Theory, pages 194--281. MIT Press, Cambridge, MA, USA, 1986.

\bibitem{Hinton2009}
Geoffrey~E Hinton and Ruslan~R Salakhutdinov.
\newblock Replicated softmax: an undirected topic model.
\newblock In {\em Advances in neural information processing systems}, pages
  1607--1614, 2009.

\bibitem{Salakhutdinov2007}
Ruslan Salakhutdinov, Andriy Mnih, and Geoffrey Hinton.
\newblock Restricted boltzmann machines for collaborative filtering.
\newblock In {\em Proceedings of the 24th International Conference on Machine
  Learning}, ICML '07, pages 791--798, New York, NY, USA, 2007. ACM.

\bibitem{Hanel20112}
R.~Hanel and S.~Thurner.
\newblock When do generalized entropies apply? {How} phase space volume
  determines entropy.
\newblock {\em EPL (Europhysics Letters)}, 96(5):50003, 2011.

\bibitem{Siwei2009}
Siwei Lyu and Eero~P Simoncelli.
\newblock Reducing statistical dependencies in natural signals using radial
  gaussianization.
\newblock In {\em Advances in neural information processing systems}, pages
  1009--1016, 2009.

\bibitem{Ramsay1998}
J.~O. Ramsay.
\newblock Estimating smooth monotone functions.
\newblock {\em Journal of the Royal Statistical Society: Series B (Statistical
  Methodology)}, 60(2):365--375, 1998.

\bibitem{Wang2001}
Fugao Wang and D.~P. Landau.
\newblock Efficient, multiple-range random walk algorithm to calculate the
  density of states.
\newblock {\em Phys. Rev. Lett.}, 86:2050--2053, Mar 2001.

\bibitem{Belardinelli2007}
R.~E. Belardinelli and V.~D. Pereyra.
\newblock Fast algorithm to calculate density of states.
\newblock {\em Phys. Rev. E}, 75:046701, Apr 2007.

\bibitem{Schneidman2006}
Elad Schneidman, Michael~J. Berry, II, Ronen Segev, and William Bialek.
\newblock {Weak pairwise correlations imply strongly correlated network states
  in a neural population.}
\newblock {\em Nature}, 440(7087):1007--12, 2006.

\bibitem{Mora2011}
Thierry Mora and William Bialek.
\newblock {Are Biological Systems Poised at Criticality?}
\newblock {\em Journal of Statistical Physics}, 144(2):268--302, 2011.

\bibitem{Stephens2013}
Greg~J. Stephens, Thierry Mora, Ga\v{s}per Tka\v{c}ik, and William Bialek.
\newblock {Statistical Thermodynamics of Natural Images}.
\newblock {\em Physical Review Letters}, 110(1):018701, 2013.

\bibitem{Mora2010}
Thierry Mora, Aleksandra~M. Walczak, William Bialek, and Curtis~G. Callan.
\newblock {Maximum entropy models for antibody diversity.}
\newblock {\em Proceedings of the National Academy of Sciences of the United
  States of America}, 107(12):5405--5410, 2010.

\bibitem{Tkacik2015}
Gašper Tkačik, Thierry Mora, Olivier Marre, Dario Amodei, Stephanie~E.
  Palmer, Michael~J. Berry, II, and William Bialek.
\newblock Thermodynamics and signatures of criticality in a network of neurons.
\newblock {\em Proceedings of the National Academy of Sciences},
  112(37):11508--11513, 2015.

\bibitem{Mora2015}
Thierry Mora, St\'{e}phane Deny, and Olivier Marre.
\newblock {Dynamical criticality in the collective activity of a population of
  retinal neurons}.
\newblock {\em Phys. Rev. Lett.}, 114(7):078105, 2015.

\bibitem{Schwab2014}
David~J. Schwab, Ilya Nemenman, and Pankaj Mehta.
\newblock {Zipf’s Law and Criticality in Multivariate Data without
  Fine-Tuning}.
\newblock {\em Phys. Rev. Lett.}, 113(6):068102, 2014.

\bibitem{Aitchison2014}
Laurence Aitchison, Nicola Corradi, and Peter~E Latham.
\newblock Zipf's law arises naturally in structured, high-dimensional data.
\newblock {\em arXiv preprint arXiv:1407.7135}, 2014.

\bibitem{Widder1946}
D.V. Widder.
\newblock {\em The Laplace transform}.
\newblock Princeton mathematical series. Princeton university press, 1946.

\bibitem{Tkacik2011}
Gašper Tkačik, Patrick Garrigan, Charles Ratliff, Grega Milčinski,
  Jennifer~M. Klein, Lucia~H. Seyfarth, Peter Sterling, David~H. Brainard, and
  Vijay Balasubramanian.
\newblock Natural images from the birthplace of the human eye.
\newblock {\em PLoS ONE}, 6(6):e20409, 06 2011.

\end{thebibliography}
\end{document}